\documentclass{conm-p-l}
\newtheorem{theorem}{Theorem}[section]
\newtheorem{proposition}[theorem]{Proposition}

\newtheorem{definition}[theorem]{Definition}

\newtheorem{remark}[theorem]{Remark}

\numberwithin{equation}{section}

\usepackage{enumerate}
\usepackage[pdftex]{graphicx}
\usepackage{graphicx}
\usepackage{amsmath}
\usepackage{euscript}
\usepackage{enumerate}
\usepackage{amsthm}

\begin{document}

\title[LAPLACIANS ON FRACTALS AND COMPLEX DYNAMICS]{The Decimation Method for Laplacians on Fractals: Spectra and Complex Dynamics}

\author{Nishu Lal }
\address{Department of Mathematics, Pomona College, Claremont, CA 91711}
\curraddr{}
\email{Nishu.Lal@pomona.edu}
\thanks{}

\author{Michel\,L.\,Lapidus}
\address{Department of Mathematics, University of California, Riverside, CA 92521-0135}
\email{lapidus@math.ucr.edu}
\thanks{The work of M.\,L.\,Lapidus was partially supported by the US National Science Foundation under the research grant DMS-1107750, as well as by the Institut des Hautes Etudes Scientifiques (IHES) where the second author was a visiting professor in the Spring of 2012 while part of this paper was written.}

\date{August the 4th, 2012.}

\dedicatory{}


\begin{abstract}
In this survey article, we investigate the spectral properties of fractal differential operators on self-similar fractals.  In particular, we discuss the decimation method, which introduces a renormalization map whose dynamics describes the spectrum of the operator.  In the case of the bounded Sierpinski gasket, the renormalization map is a polynomial of one variable on the complex plane.  The decimation method has been generalized by C. Sabot to other fractals with blow-ups and the resulting associated renormalization map is then a multi-variable rational function on a complex projective space.  Furthermore, the dynamics associated with the iteration of the renormalization map plays a key role in obtaining a suitable factorization of the spectral zeta function of fractal differential operators.  In this context, we discuss the works of A. Teplyaev and of the authors regarding the examples of the bounded and unbounded Sierpinski gaskets as well as of fractal Sturm--Liouville differential operators on the half-line.
\end{abstract}
\maketitle
\newpage \setcounter{tocdepth}{2}
\tableofcontents

\section{Introduction}
From the probabilistic point of view, the Laplacian on the Sierpinski gasket $SG$ was introduced independently by S. Goldstein in \cite{Go} and S. Kusuoka in \cite{Ku87} (and a little later, by M. Barlow and E. Perkins in \cite{BP98}), as the generator of the semigroup associated with Brownian motion on $SG$.  (See, e.g., \cite{B91} and \cite{B98} for early reviews of the subject of diffusions and random walks on self-similar fractals.)  However, from the point of view of analysis, which will be our main concern here, the Laplace operator was first defined by J. Kigami \cite{Kigami89} for the Sierpinski gasket and was later extended in \cite{Kigami93} to a class of self-similar fractals, called the \emph{post critically finite} sets (p.c.f. sets).  (See, e.g., \cite{Kigami01} and \cite{RS06} for a detailed exposition.)  The Laplacian on a p.c.f. set is defined as the limit of a sequence of Laplacians of finite graphs that approximate the fractal.  Following the work of the physicists R. Rammal and G. Toulouse \cite{RR84, RT83}, M. Fukushima and T. Shima \cite{FS92, TS96} studied the eigenvalue problem associated with the Laplacian on the Sierpinski gasket and introduced the \emph{decimation method} in order to give an explicit construction of the set of eigenvalues.  The decimation method, described in \S2 of the present paper, is a process through which we find the spectrum of the Laplacian on a fractal (in a certain class of self-similar sets) via the iteration of a rational function of a single complex variable, called the \emph{renormalization map}.  In the case of the finite (or bounded) Sierpinski gasket, this rational map is a polynomial on the complex plane and its dynamics is not too difficult to understand in order to analyze the spectrum of the Laplacian.

Later on, C. Sabot (\cite{CS98}--\cite{CS05}) generalized the decimation method to Laplacians defined on a class of finitely-ramified self-similar sets with blow-ups.  First, in \cite{CS98, CS01, CS05}, he studied fractal Sturm--Liouville operators on the half-line, viewed as a blow-up of the self-similar unit interval, and discovered that the corresponding decimation method then involves the dynamics of a rational map which is no longer a function of a single complex variable but is instead defined on the two-dimensional complex projective space; see \S3.  (It therefore arises from a homogeneous rational function of three complex variables.)  This rational map is initially defined on a space of quadratic forms associated with the fractal and its construction involves the notion of trace of a symmetric matrix on a (finite) subset of this space; see \S4.  From the point of view of multi-variable complex dynamics, the set of eigenvalues (or spectrum) is best understood in terms of an invariant curve under the iteration of the rational map.

Using as a model the unbounded (or infinite) Sierpinski gasket (the so-called Sierpinski lattice, see \S4 and Figure \ref{fig:UnboundedSG}), Sabot \cite{CS} used Grassmann algebras in order to construct the renormalization map for other lattices based on (symmetric) finitely ramified self-similar sets.  The idea is that one embeds the space of symmetric matrices in a Grassmann algebra in order to analyze the operation of taking the trace on a (suitable) finite subset.  In some sense, this enables one to linearize this operation.  The polynomial associated with the classical bounded Sierpinski gasket, initially introduced in the work of Rammal and Toulouse \cite{RR84, RT83} and later rigorously formalized in \cite{FS92, TS96}, can then be recovered from the renormalization map associated with the unbounded Sierpisnki gasket.  

Finally, in recent work, Lal and Lapidus \cite{LL12} have studied the spectral zeta function of the Laplacian on a suitable self-similar set and established a factorization formula for the associated spectral zeta function in terms of a certain hyperfunction, a geometric zeta function and a zeta function associated with the iteration of a renormalization map, which is a multi-variable rational map acting in a complex projective space.  This latter work in \cite{LL12} extends to several complex variables an earlier factorization formula due to Teplyaev \cite{AT04, AT07}, itself extending the second author's factorization formula \cite{Lapidus92, Lapidus93} for the spectral zeta function of a fractal string (see also \cite{LM95, LP93, LvanF06} for various applications of, and motivations for, the latter factorization).  We survey some of these results in the last part of this paper; see \S5.

In closing this introduction, we mention that the work of \cite{LL12} described in \S5 focuses on two different models, namely, fractal Sturm--Liouville differential operators on the half-line (as in \S3) and the infinite (or unbounded) Sierpinski gasket (as in \S4).  In each of these cases, the Dirac delta hyperfunction plays a key role in the rigorous mathematical formulation of the factorization results.

\section{The bounded Sierpinski gasket}
The (bounded or finite) Sierpinski gasket (SG) is generated by the iterated function system (IFS) consisting of three contraction mappings $\Phi_j:\mathbb{R}^2 \rightarrow \mathbb{R}^2$ defined by 
\begin{equation}
\label{SGContractions}
\Phi_j(x)=\frac{1}{2}(x - q_j)+q_j
\end{equation}
for $j=0,1,2$, where $q_0,q_1,q_2$ are the vertices of an equilateral triangle.  (See Figure 1.)  Note that each $\Phi_j$ has a unique fixed point, namely, $q_j$.  The Sierpinski gasket is the unique (nonempty) compact subset of $\mathbb{R}^2$ such that $SG=\Phi_1(SG) \cup \Phi_2(SG) \cup \Phi_3(SG)$.  

\begin{figure}
  \centering
    \includegraphics*[scale=0.37]{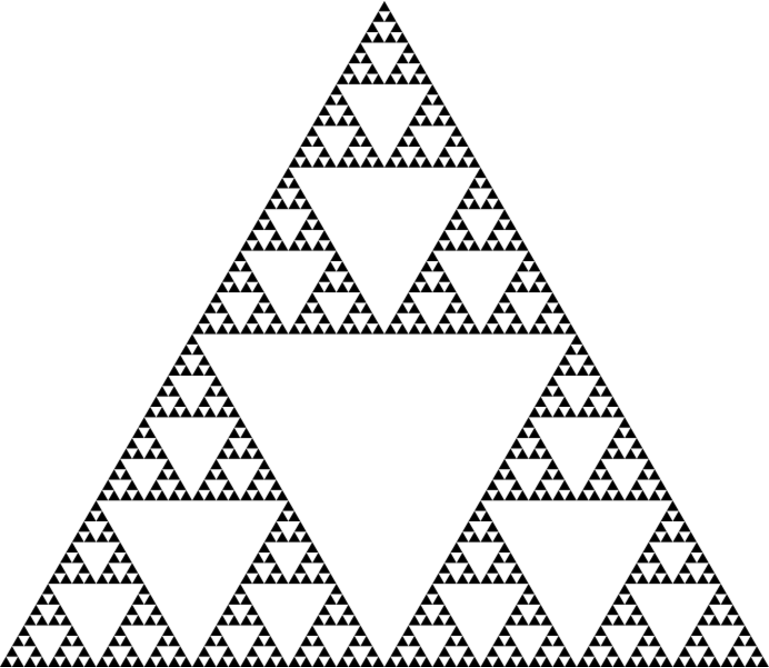}
  \caption{The (bounded) Sierpinski gasket $SG$.
\label{fig:sierpinskigasket}}
\end{figure}

For each integer $m\geq0$, denote by $\Gamma_m$ the mth level finite graph approximating $SG$, and let $\Gamma_0$ be the complete graph on $V_0:=\{q_1, q_2, q_3\}$.
The set $V_m$ of vertices of $\Gamma_m$ is defined recursively as follows: 
\[
V_m:=\bigcup_{j=0}^2\Phi_j(V_{m-1}), 
\]
for each integer $m \geq 1$.

Let $\ell^2(V_m)$ be the space of real-valued functions on $V_m$, equipped with the standard inner product $(u, v)=\sum_{x \in V_m} u(x)v(x)$.
The discrete Laplacian on $\ell^2(V_m)$ (or the finite graph Laplacian on $\Gamma_m$) is defined by 
\[
\Delta_m f(x)=\frac{1}{4} \sum_{y \sim x} f(x)-f(y),
\]
where $x \in \Gamma_m$ and the sum is extended over all neighbors $y$ of $x$ in the graph $\Gamma_m$. 

We then define the Laplacian $\Delta=\Delta_{\mu}$ on SG as the following limit of rescaled finite-difference operators:
\[
\Delta f(x)=\lim_{m\rightarrow \infty} 5^m \Delta_m f(x).
\]
The factor 5 is the product of the scaling factor 3 for the natural Haursdorff measure on $SG$ and the renormalized factor $\frac{5}{3}$  for the energy.  Here and thereafter, when we write $\Delta=\Delta_{\mu}$, the subscript $\mu$ refers to the natural self-similar (or equivalently, in this case, Hausdorff) probability measure on $SG$.  

The graph energy on each $V_m$ ($m\geq 0$) is defined by 
\[
\mathcal{E}_m (u, u)= \bigg(\frac{3}{5}\bigg)^{-m} \sum_{x \sim y}(u(x)- u(y))^2, 
\]
which does not change under the process of harmonic extension, and the graph energy on SG is then defined by 
\[
\mathcal{E}(u, u)=\sup_{m\geq 0} \mathcal{E}_m(u, u)=\lim_{m\rightarrow \infty} \mathcal{E}_m(u, u) 
\]
(this limit always exists in $[0, \infty]$ since the sequence $\{\mathcal{E}_m(u, u)\}_{m=0}^{\infty}$ is nondecreasing).  In the sequel, we also write $\mathcal{E}(u)=\mathcal{E}(u, u)$, in short, to refer to this quadratic form.  (By definition, $u$ belongs to the domain of $\mathcal{E}$ if and only if $\mathcal{E}(u) < \infty$.)  The associated bilinear form $\mathcal{E}(u, v)$ is then defined by polarization.

Suppose $u$ is a function defined on $V_0$, with values at each of the three vertices of $V_0$ denoted by $a$, $b$, and $c$; see Figure 2.  We want to extend $u$ to $V_1$ in such a way that it minimizes the energy. 

Let $\tilde{u}$ be the harmonic extension of $u$ to $V_1$ and denote by $x$, $y$ and $z$ the values of $\tilde{u}$ to be determined at each of the three vertices of $V_1\setminus V_0$; see Figure 2.  Since, by definition, $\tilde{u}$ minimizes $\mathcal{E}_1(v)$ subject to the constraint $v=u$ on $V_0$, we can take the partial derivatives with respect to $x, y, z$ and set them equal to zero to obtain

\[
\begin{aligned}
4x&= b+c+y+z\\
4y&= a+c+x+z\\
4z&= a+b+x+y.\\
\end{aligned}
\]

These equations express the mean value property of a (discrete) harmonic function, according to which the function value at each of the junction points is the average of the function values of the four neighboring points in the graph.  
We can use the matrix representation of these equations\\

\begin{center}
$\left( \begin{array}{ccc}
4 &  -1 & -1 \\
-1 & 4 & -1   \\
-1 &-1 & 4 \end{array} \right)$  
$\left( \begin{array}{ccc}
x \\
y   \\
z \end{array} \right)$  
=$\left( \begin{array}{ccc}
b+c \\
a+c \\
a+b\end{array} \right)$ 
\end{center}
\vspace{1pc}
\noindent in order to obtain the following solutions:\\
\[
\begin{aligned}
x&=\frac{1}{5}a +\frac{2}{5}b+\frac{2}{5}c\\
y&=\frac{2}{5}a +\frac{1}{5}b+\frac{2}{5}c\\
z&=\frac{2}{5}a +\frac{2}{5}b+\frac{1}{5}c.\\
\end{aligned}
\]
\vspace{1pc}

\begin{figure}
  \centering
  \includegraphics*[scale=0.37]{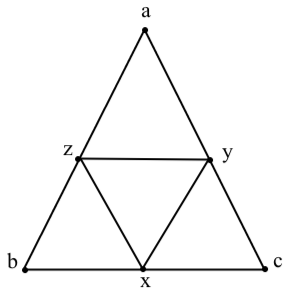}
 \caption{The values of the harmonic extension $\tilde{u}$ at each vertex in $V_1$.  The values $a, b, c$ at each vertex in $V_0$ are prescribed, whereas the values $x, y, z$ at each vertex in $V_1\setminus V_0$ are uniquely determined by requiring that the energy be minimized in the passage from $V_0$ to $V_1$.} 
 \end{figure}

The harmonic extension $\tilde{u}$ therefore satisfies the $\frac{1}{5}-\frac{2}{5}$ rule.  See \cite{Kigami89}, \cite{Kigami01}, \cite{RS06}.  (More generally, this rule also holds for the harmonic extension on each m-cell of $SG$.)  More specifically, iterating this process to each finite graph $V_m$, one then defines $\tilde{u}$ on the countable set of vertices $V^*:=\cup_{m\geq0} V_m$ and finally, extends it to all of $SG$, by continuity and using the density of $V^*$ in $SG$.  The resulting function, still denoted by $\tilde{u}$, is called the \emph{harmonic extension} of $u$.  According to Definition \ref{HarmonicExtension} below (which is also a theorem), it is the unique harmonic function (i.e., $\Delta \tilde{u}=0$) such that $\tilde{u}=u$ on $V_0$.  Equivalently, it is the unique minimizer of the energy functional $\mathcal{E}=\mathcal{E}(v)$ subject to the constraint $v=u$ on $V_0$.

\begin{definition}
\label{HarmonicExtension}
A harmonic function on SG \emph(with boundary value $u$ on $V_0$\emph) is a continuous function whose restriction to any $\Gamma_m$ or $V_m$ is the harmonic extension of $u$.  In other words, it is the unique solution of the following Poisson problem\emph: $\Delta v = 0$, $v=u$ on $V_0$.  It must therefore necessarily coincide with the harmonic extension $\tilde{u}$ of $u$ to all of $SG$.
\end{definition}

\subsection{Spectral properties of the Laplacian on the Sierpinski gasket}
The Laplacian operators on p.c.f. self-similar fractals are defined similarly via a suitable approximation.  To study the spectrum of the Laplacian, we consider the equation $-\Delta u =\lambda u$, where u is a continuous function.  The spectrum of the Laplacian on the Sierpinski gasket was first studied in detail by the physicists R. Rammal and G. Toulouse \cite{RR84, RT83}. Later on, M. Fukushima and T. Shima \cite{FS92, TS96} gave a precise mathematical description of the eigenvalues and the eigenfunctions.  Still in the case of the Sierpinski gasket, Rammal and Toulouse discovered interesting relations between the spectrum of the discrete Laplace operator and the iteration of a polynomial of one complex variable, $R=R(z):=z(5-4z)$.  More precisely, for any $m\geq 0$, if $\lambda$ is an eigenvalue of $-\Delta_{m+1}$ on $\Gamma_{m+1}$, then $\lambda(5-4\lambda)$ is an eigenvalue of $-\Delta_m$ on $\Gamma_m$.  Thus, the relationship between the eigenvalues of the Laplacians on one graph and it successor can be described by a quadratic equation, $\lambda_m= \lambda_{m+1}(5-4\lambda_{m+1})=R(\lambda_{m+1})$.  The restriction to $V_m$ of any eigenfunction belonging to $\lambda_{m+1}$  is an eigenfunction belonging to $\lambda_{m}$.  The relationship between the eigenvalues $\lambda_m$ and $\lambda_{m+1}$ of $-\Delta_{m}$ and $-\Delta_{m+1}$, respectively, can be found by comparing the corresponding eigenvalue problem for a point common to both $V_{m}$ and $V_{m+1}$.

\begin{theorem} [Fukushima and Shima, \text{\cite{FS92}, \cite{TS96}}]
\label{FuSh}
\hspace{1mm}
\begin{enumerate}[\emph(i\emph)]
\item If $u$ is an eigenfunction of $-\Delta_{m+1}$ with eigenvalue $\lambda$ \emph(that is, $-\Delta_{m+1}u = \lambda u$\emph), and if $\lambda \notin B$, then $-\Delta_m (u|_{V_m}) = R(\lambda)u|_{V_m}$, where $B:= \{\frac{1}{2}, \frac{5}{4}, \frac{3}{2}\}$ is the set of `forbidden' eigenvalues and $u|_{V_m}$ is the restriction of $u$ to $V_m$.
\item If $-\Delta_m u = R(\lambda) u$ and $\lambda\notin B$, then there exists a unique extension $w$ of $u$ to $V_{m+1}$ such that
$-\Delta_{m+1}w= \lambda w$.
\end{enumerate}
\end{theorem}

\begin{figure}
\centering

\includegraphics[width=5in,height=3.5in]{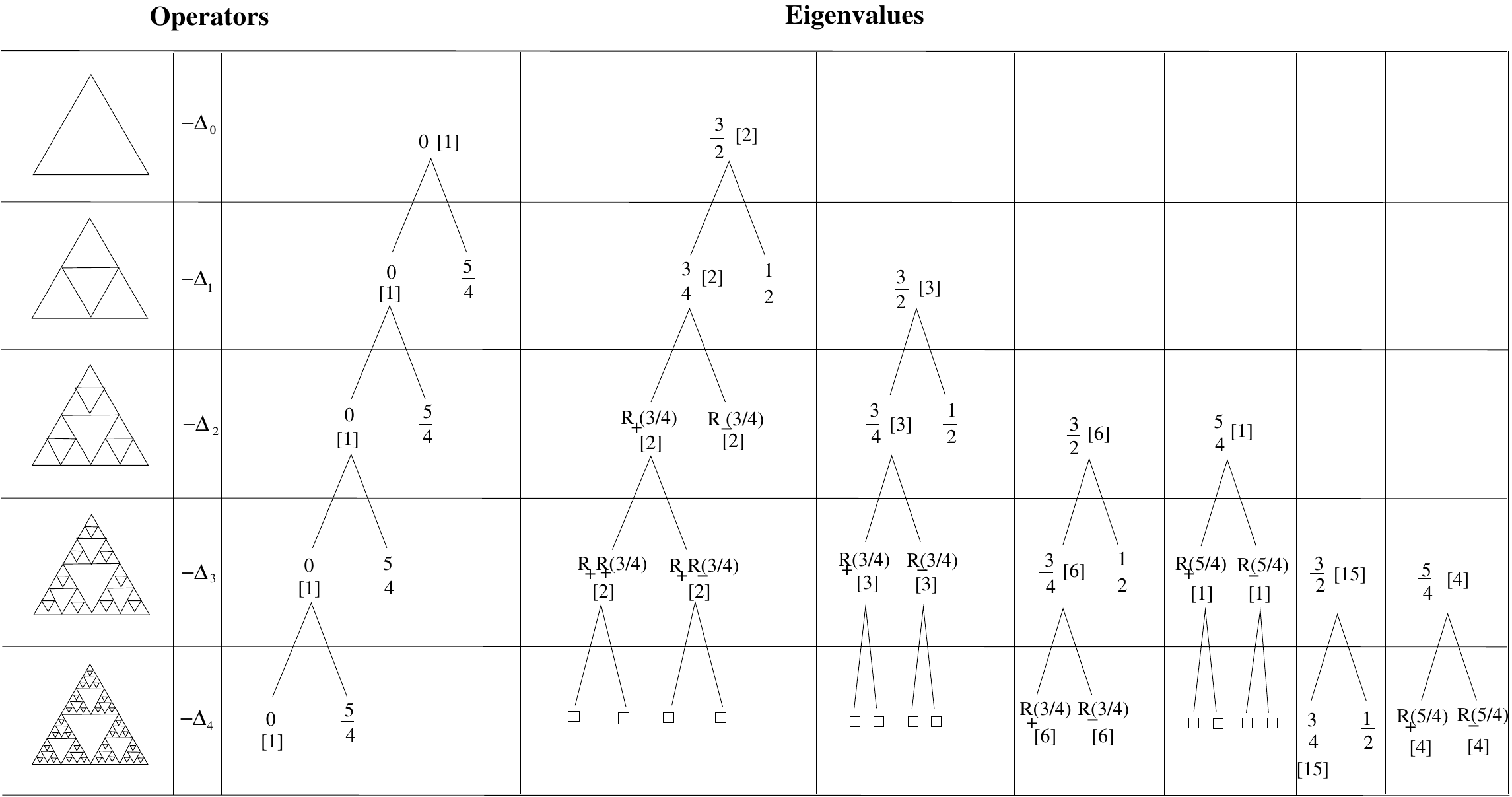}

\caption{The eigenvalues of $SG$ are obtained via the decimation method as limits of inverse images of the renormalization map $R$.  At each level, the two branches of the tree correspond to the two inverse branches of the quadratic polynomial $R$, denoted by $R_+$ and $R_-$.  The number in the bracket represents the multiplicity of the corresponding eigenvalue.
\label{fig:Nishuu3}}
\end{figure}

At any given level $m$, there are two kind of eigenvalues of $-\Delta_m$, called the initial and continued eigenvalues.  The continued eigenvalues arise from the spectrum of $-\Delta_{m-1}$ via the decimation method (described in Theorem \ref{FuSh}), and the remaining eigenvalues are called the initial eigenvalues. The forbidden eigenvalues $\{\frac{1}{2}, \frac{5}{4}, \frac{3}{2}\}$ in Theorem \ref{FuSh} have no predecessor, i.e., they are the initial eigenvalues.  Furthermore, the exclusion of the eigenvalue $\frac{1}{2}$ can be explained by showing that $\frac{1}{2}$ is an eigenvalue of $-\Delta_m$ only for $m=1$.  (See Figure \ref{fig:Nishuu3}.)

Given that the eigenvalues of $-\Delta_0$ are $\{0, \frac{3}{2} \}$, we consider the inverse images of $0$ and $\frac{3}{2}$ under $R$ (that is, their images under $R_{-}(z) = \frac{5-\sqrt{25-16z}}{8}$ and $R_{+}(z) = \frac{5+\sqrt{25-16z}}{8}$, the two inverse branches of the quadratic polynomial $R(z)=z(5-4z)$), to obtain the eigenvalues of $-\Delta_1$.  The continuation of this process generates the entire set of eigenvalues for each level.  The diagram provided in Figure \ref{fig:Nishuu3} illustrates the eigenvalues associated with each graph Laplacian $-\Delta_m$, for $m=0, 1, 2,...$, in terms of the inverse iterates of the polynomial $R$.

The spectrum of the Sierpinski gasket is the renormalized limit of the spectra of the graph Laplacians $-\Delta_m$.  More specifically, each eigenvalue satisfying the equation $-\Delta_{\mu} u=\lambda u$ can be written as 

\begin{equation}
\label{EigenvaluesSG}
\lambda= \lim_{m \rightarrow \infty} 5^m \lambda_m,
\end{equation}
for a sequence $\{\lambda_m\}^{\infty}_{m=m_0}$ such that $\lambda_m=\lambda_{m+1}(5-4\lambda_{m+1})=R(\lambda_m)$ for all $m \geq m_0$ and for some smallest integer $m_0$ (which is allowed to depend on $\lambda$).  Note that for $m > m_0$, $\lambda_m$ does not coincide with any of the forbidden eigenvalues in $\{\frac{1}{2}, \frac{5}{4}, \frac{3}{4}\}$, whereas $\lambda_{m_0}$ belongs to the set $\{\frac{1}{2}, \frac{5}{4}, \frac{3}{4}\}$.  Furthermore, the values $\lambda_m$ are determined by the solutions of $\lambda_m=\lambda_{m+1}(5-4\lambda_{m+1})$:
$\lambda_{m+1}=\frac{5+\epsilon_m\sqrt{25-16\lambda_m}}{8}$, where $\epsilon_m=\pm1$, provided that the limit in \eqref{EigenvaluesSG} exists.  The limit $\lambda$ only exists if $\epsilon_m=-1$ for all but finitely many integers $m$.  In that case, $\lambda$ is an eigenvalue of $-\Delta_{\mu}$ where, as before, $\Delta=\Delta_{\mu}$ denotes the Laplacian on $SG$.  And conversely, every eigenvalue of $-\Delta_{\mu}$ can be obtained in this manner.

In the next section, we will discuss a generalization (due to Sabot \cite{CS01, CS03}) of the decimation method  to rational functions of several complex variables, when presenting the case of fractal Sturm--Liouville differential operators.  The extended (multi-variable) decimation method is not valid for all self-similar fractals, in general.  However, it does apply to a large class of symmetric p.c.f. self-similar fractals (whereas the original single-variable decimation method only applied to a rather restricted and difficult to characterize class of symmetric finitely ramified and hence, p.c.f., self-similar fractals).  Therefore, this generalization is a very significant extension of the original (single-variable) decimation method, for which it also provides a nice geometric and algebraic interpretation (see \S4).

In fact, mathematically, the truly beautiful underlying structure of the extended decimation method is only revealed by considering the multi-variable case, even in the original setting of a single complex variable.  (The latter one-variable case should really be formulated in terms of two complex variables or equivalently, on the complex projective line, and via phase space symplectic geometry or Grassmann algebras, in terms of two conjugate variables.  In hindsight, the reduction to a single variable is simply a confusing, albeit convenient, artifact in this situation.)

\begin{remark}
The interested reader can find in \cite{BCD08, BC08} some detailed and relatively elementary computations pertaining to the decimation method and the associated renormalization map \emph(viewed only as a rational function of one complex variable\emph) in the case of certain examples of finitely ramified, symmetric self-similar fractals.
\end{remark}

\section{Generalization of the decimation method}

\subsection{The fractal Sturm--Liouville operator}
C. Sabot, in a series of papers (\cite{CS98}--\cite{CS05}), extended the decimation method to Laplacians defined on a class of (symmetric) finitely ramified (really, p.c.f.) self-similar sets with blow-ups.  This extension involves the dynamics of rational functions of several complex variables.   We discuss the prototypical example he studied, fractal Laplacians on the blow-up $I_{<\infty>}=[0, \infty)$ of the unit interval $I=I_{<0>}=[0,1]$.  From now on, we will assume that 
\begin{equation}
\label{constants}
0<\alpha<1 \mbox{,  } b=1-\alpha, \mbox{  } \delta =\frac{\alpha}{1-\alpha}, \mbox{ and } \gamma=\frac{1}{\alpha(1-\alpha)}. 
\end{equation}
Consider the contraction mappings from $I=[0, 1]$ to itself given by
\[
\Psi_1(x)= \alpha x, \mbox{  } \Psi_2(x)=1-(1-\alpha)(1-x),
\]
and the unique self-similar measure $m$ on $[0, 1]$ such that for all $f \in C([0, 1])$ (the space of continous functions on $I=[0, 1]$), 
\begin{equation}
\label{IntegralIdentity}
\int_0^1 f dm = b\int_0^1 f\circ\Psi_1 dm + (1-b)\int_0^1 f\circ\Psi_2 dm.
\end{equation}
Here, $I$ is viewed as the unique \emph{self-similar set} (in the sense of \cite{Hu81}) associated with the iterated function system $\{\Psi_1, \Psi_2\}$:

\begin{equation}
\label{Self-similarityEq}
I=\Psi_1(I) \cup \Psi_2(I).
\end{equation}

Define $H_{<0>} = -\frac{d}{dm}\frac{d}{dx}$, the free Hamiltonian with Dirichlet boundary conditions on $[0, 1]$, by $H_{<0>} f= g$ on the domain 
\[
\bigg\{ f\in L^2(I, m): \exists g \in L^2(I, m), f(x)= cx + d + \int_0^x \int_0^y g(z)dm(z) dy, f(0)= f(1)=0 \bigg\}.  
\]
The operator $H_{<0>}$  is the infinitesimal generator associated with the Dirichlet form $(a,\mathcal{D})$ given by 
\[
a(f,g) = \int_0^1 f^{\prime} g^{\prime}dx, \mbox{ for } f,g \in \mathcal{D},  
\]
where
\[
\mathcal{D} = \{ f \in L^2(I, m): f^{\prime} \in L^2(I, dx)\}.
 \]

\noindent As can be easily checked, the Dirichlet form $a$ satisfies the similarity equation
\begin{equation}
\label{SelfsimilarIdentity}
a(f)=\alpha^{-1} a(f \circ \Psi_1) + (1-\alpha)^{-1} a(f \circ \Psi_2),
\end{equation}
where we denote the quadratic form $a(f, f)$ by $a(f)$.
(See, e.g., \cite{UF03} for an exposition.)

The idea is that viewing the unit interval $I$ as a self-similar set, as in \eqref{Self-similarityEq} above, we construct an increasing sequence of intervals $I_{<n>}$, for $n=0, 1, 2, ...$, by blowing-up the initial unit interval by the scaling ratio $\alpha^{-n}$.  Hence, we can extend all the objects involved, ($m$, $a$, $H_{<0>}$), from $I= I_{<0>}$ to $I_{<n>} = \Psi_1^{-n}(I)=[0,\alpha^{-n}]$, which can be expressed as a self-similar set as follows:
\[
I_{<n>} = \bigcup_{i_1,...,i_n}\Psi_{i_1...i_n}(I_{<n>}), 
\]
where $(i_1,...,i_n) \in {\{1,2\}}^{n}$.  Here, we have set $\Psi_{i_1...i_n}=\Psi_{i_n} \circ ...\circ \Psi_{i_1}$.  

More precisely, for each $n\geq 0$, we define the self-similar measure $m_{<n>}$ on $I_{<n>}$ by 
\[
\int_{I_{<n>}}fdm_{<n>} = (1-\alpha)^{-n} \int_I f\circ\Psi_1^{-n}dm, 
\]
for all $f \in C(I_{<n>})$.  Similarly, the corresponding differential operator, $H_{<n>}= -\frac{d}{dm_{<n>}}\frac{d}{dx}$ on $I_{<n>}=[0, \alpha^{-n}]$, can be defined  as the infinitesimal generator of the Dirichlet form
$(a_{<n>},\mathcal{D}_{<n>})$ given by 
\[
a_{<n>}(f) = \int_0^{{\alpha}^{-n}} (f')^2 dx = \alpha^n a(f\circ\Psi_1^{-n}), \mbox{ for }  f \in \mathcal{D}_{<n>},
\]
where
\[
\mathcal{D}_{<n>} = \{f \in L^2(I_{<n>}, m_{<n>}):  f' \mbox{ exists a.e. and } f' \in L^2(I_{<n>}, dx)\}.
\]

We define $H_{<\infty>}$ as the operator $ -\frac{d}{dm_{<\infty>}}\frac{d}{dx}$ with Dirichlet boundary conditions on $I_{<\infty>}= [0, \infty)$.  It is clear that the (projective system of) measures $m_{<n>}$ give rise to a measure $m_{<\infty>}$ on $I_{<\infty>}$ since for any $f \in \mathcal{D}_{<n>}$ with support contained in $[0, 1]$, $a_{<n>}(f)=a(f)$ and $\int_{I_{<n>}} fdm_{<n>} =\int_{I} f dm$.  Furthermore, we define the corresponding Dirichlet form $(a_{<\infty>}, \mathcal{D}_{<\infty>})$ by 
\[
a_{<\infty>}(f)=\sup_{n\geq 0}a_{<n>}(f|_{I_{<n>}})=\lim_{n \rightarrow \infty} a_{<n>}(f|_{I_{<n>}}), \mbox{ for } f \in \mathcal{D}_{<\infty>},
\]
where 
\[
\mathcal{D}_{<\infty>}=  \{f \in L^2(I_{<\infty>}, m_{<\infty>}): \sup_n  a_{<n>}(f|_{I_{<n>}}) < \infty\}.
\]
Clearly, $a_{<\infty>}$ satisfies a self-similar identity analogous to Equation \eqref{SelfsimilarIdentity} and its infinitesimal generator is $H_{<\infty>}$.

\subsection{The eigenvalue problem}
The study of the eigenvalue problem 
\begin{equation}
\label{EigenProb}
H_{<n>}f = -\frac{d}{dm_{<n>}}\frac{d}{dx}f = \lambda f
\end{equation}
for the Sturm--Liouville operator $H_{<n>}$, equipped with Dirichlet boundary conditions on $I_{<n>}$, revolves around a map $\rho$, called the \textit{renormalization map}, which is initially defined on a space of quadratic forms associated with the self-similar set $I_{<n>}$ (or with a corresponding finite graph, see Remark \ref{Remark3.1} below) and then, via analytic continuation, on $\mathbb{C}^3$ as well as (by homogeneity) on $\mathbb{P}^2(\mathbb{C})$, the complex projective plane.  (More precisely, the renormalization map is associated to the passage from $I_{<n-1>}$ to $I_{<n>}$ or, equivalently, from $I_{<0>}$ to $I_{<1>}$.)  As will be explained in \S3.3, the propagator of the above differential equation \eqref{EigenProb} is very useful in producing this rational map, initially defined on $\mathbb{C}^3$, and later on, as the polynomial map
\begin{equation}
\label{RenormalizationMap}
\rho([x,y,z])= [x(x+\delta^{-1}y)-\delta^{-1}z^2, \delta y(x+\delta^{-1}y)-\delta z^2, z^2],
\end{equation}
defined on the complex projective plane $\mathbb{P}^2(\mathbb{C})$.  Here, $[x, y, z]$ denotes the homogeneous coordinates of a point in $\mathbb{P}^2(\mathbb{C})$, where $(x, y, z) \in \mathbb{C}^3$ is identified with $(\beta x, \beta y, \beta z)$ for any $\beta \in \mathbb{C}$, $\beta \neq 0$. Clearly, in the present case, $\rho$ is a homogeneous quadratic polynomial.  

(Note that if $\rho$ is viewed as a map from $\mathbb{P}^2(\mathbb{C})$ to itself, then one should write more correctly, 
\[
\rho([x: y: z])=[x(x+\delta^{-1} y)-\delta^{-1}z^2: \delta y(x+\delta^{-1} y)-\delta z^2: z^2],
\]
where $[u: v: w]=[u, v, w]$ denote the homogeneous coordinates of a generic point in the projective plane $\mathbb{P}^2=\mathbb{P}^2(\mathbb{C})$.  We will use this notation in \S4.)

As we shall see in \S3.3, the spectrum of the fractal Sturm--Liouville operator $H_{<\infty>}$, as well as of its finite graph (or rather, here, bounded interval, see Remark \ref{Remark3.1} below) approximations $H_{<n>}$ ($n=0, 1, 2, ...$), is intimately related to the iteration of $\rho$.  The spectrum of $H_{<0>}$ and of $H_{<n>}$ ($n=1, 2, ...$) is discrete for any value of $\alpha$ in $(0, 1)$ (and hence, for any $\gamma \geq4$).  (See, e.g., \cite{BNST03}, \cite{UFF03}, \cite{UF03},  \cite{UF05}, \cite{UFMZ02}.) However, in the sequel, we will focus our attention on the case where $\alpha \leq \frac{1}{2}$ (or equivalently, $\delta \leq 1$).  In that case, the spectrum of $H_{<\infty>}$ is pure point for $\alpha < \frac{1}{2}$ (i.e., $\gamma > 4$), but absolutely continuous for $\alpha=\frac{1}{2}$ (i.e., for $\gamma=4$).  Furthermore, observe that since $\alpha \in (0, 1)$ and $\gamma=\frac{1}{\alpha(1-\alpha)}$ (see Equation \eqref{constants}), we always have $\gamma\geq 4$ (and in particular, $\gamma>1$), independently of the above assumption according to which $\alpha\leq \frac{1}{2}$.  Finally, note that $\gamma =4$ if and only if $\alpha=\frac{1}{2}$ (i.e., $\delta=1$), an interesting special case which will be investigated at the end of \S5.

\begin{remark}
\label{Remark3.1}
In \cite{CS01} \emph(see also \cite{CS03} and \cite{CS}\emph), is also provided a description of the renormalization map in terms of lattice or finite graph \emph(rather than of bounded self-similar interval\emph) approximations of the half-line.  By necessity of concision, we will not  discuss this matter further in this paper.  The method, however, is very analogous to the one described in part of \emph{\S4} below in the case of the infinite lattice $SG^{<\infty>}$ based on the bounded Sierpinski gasket $SG$.  \emph(In some definite sense, the bounded self-similar interval $I=[0, 1]$ would play the role here of $SG$, while the half-line $I_{<\infty>}=[0, \infty)$ would be a substitute for $SG^{<\infty>}$.\emph)
\end{remark}

\subsection{The renormalization map and the spectrum of the operator}
The renormalization map $\rho$ is a function from the complex projective plane $\mathbb{P}^2(\mathbb{C})$ to itself which is induced by the above homogeneous polynomial map from $\mathbb{C}^3$ to itself; see Equation \eqref{RenormalizationMap} and the discussion surrounding it.  As we shall see later on, the spectrum of the operator $H_{<\infty>}$ defined in \S3.1 is intimately connected with the iteration of the renormalization map.  Following \cite{CS01}, we next explain how the explicit expression for the renormalization map given in Equation \eqref{RenormalizationMap} above can be derived by studying the propagator for the eigenvalue problem associated with the operator.  

We define the \textit{propagator} $\Gamma_{\lambda}(s, t)$ for the eigenvalue problem $-\frac{d}{dm_{<\infty>}}\frac{d}{dx}f=\lambda f$ associated with the operator $H_{<\infty>}$ on $I_{<\infty>}=[0, \infty)$ as a time evolution function which, for each $0\leq s \leq t$, is a $2 \times 2$ matrix with nonzero determinant such that the solution of the equation satisfies 

\[
\left[ \begin{array}{c} f(t)\\ f^{\prime}(t) \end{array} \right] = \Gamma_{\lambda}(s, t)  \left[ \begin{array}{c} f(s) \\ f^{\prime}(s) \end{array} \right],
\]
where $f^\prime$ denotes the derivative of $f$.

Using the self-similarity relations \eqref{IntegralIdentity} and \eqref{SelfsimilarIdentity} satisfied by the measure $m$ and the Dirichlet form $a$, respectively, and recalling that $\gamma$ is given by Equation \eqref{constants}, we obtain $\Gamma_{<n>, \lambda}=D_{\alpha^n} \circ \Gamma_{\gamma^n\lambda} \circ D_{\alpha^{-n}}$ for the propagator $\Gamma_{<n>,\lambda}$ associated with the eigenvalue problem  $-\frac{d}{dm_{<n>}}\frac{d}{dx}f=\lambda f$, where
\[
D_{\alpha^n}:=\begin{bmatrix} 1 & 0 \\ 0 & \alpha^n \end{bmatrix}. 
\]

In order to derive the expression of the renormalization map stated in Equation \eqref{RenormalizationMap}, we will consider the case when $n=1$.  Thus, we have $\Gamma_{<1>, \lambda}=D_{\alpha} \circ \Gamma_{\gamma\lambda} \circ D_{\alpha^{-1}}$.  Let 
\[
\Gamma_{\lambda}:=\begin{bmatrix} a(\lambda) & b(\lambda) \\ c(\lambda) & d(\lambda) \end{bmatrix}.
\]
\noindent We will proceed with the following calculations:
\[
\begin{aligned}
\Gamma_{<1>, \lambda}&= 
\begin{bmatrix} 1 & 0 \\ 0 & \alpha \end{bmatrix} \begin{bmatrix} a(\gamma \lambda) & b(\gamma \lambda) \\ c(\gamma \lambda) &d(\gamma \lambda) \end{bmatrix} \begin{bmatrix} 1 & 0 \\ 0 & \alpha^{-1} \end{bmatrix}
&=\begin{bmatrix} a(\gamma \lambda) & \alpha^{-1}b(\gamma \lambda) \\ \alpha c(\gamma \lambda) & d(\gamma \lambda) \end{bmatrix}.
\end{aligned}
\]

\noindent On the other hand, we have 
\[
\begin{aligned}
\Gamma_{<1>, \lambda} &= \Gamma_{\lambda}(1, \alpha^{-1}) \circ \Gamma_{\lambda}(0, 1) = D_{\delta} \circ \Gamma_{\lambda} \circ D_{\delta^{-1}} \circ \Gamma_{\lambda}\\
&=\begin{bmatrix} 1 & 0 \\ 0 & \delta \end{bmatrix} 
\begin{bmatrix} a(\lambda) & b(\lambda) \\ c(\lambda) & d(\lambda) \end{bmatrix}
\begin{bmatrix} 1 & 0 \\ 0 & \delta^{-1} \end{bmatrix}
\begin{bmatrix} a(\lambda) & b(\lambda) \\ c(\lambda) & d(\lambda) \end{bmatrix} \\
&=\begin{bmatrix} a(\lambda)^2+\delta^{-1}b(\lambda)c(\lambda) &a(\lambda)b(\lambda)+\delta^{-1}d(\lambda)b(\lambda) \\ \delta a(\lambda) c(\lambda)+c(\lambda) d(\lambda) & \delta b(\lambda)c(\lambda)+d(\lambda)d(\lambda) \end{bmatrix}.
\end{aligned}
\]

\noindent Using the fact that $\Gamma_{\lambda} \in SL_2(\mathbb{C})$ (the special complex linear group of $2 \times 2$ matrices) and hence, that $a(\lambda)d(\lambda)-b(\lambda)c(\lambda) =1$, we see that the two diagonal entries can be rewritten as
\[
a(\lambda)^2+\delta^{-1}b(\lambda)c(\lambda)=a(\lambda) \bigg[a(\lambda)+\delta^{-1}\bigg(\frac{d(\lambda)a(\lambda)-1}{a(\lambda)}\bigg)\bigg]=a(\lambda)(a(\lambda)+\delta^{-1}d(\lambda))-\delta^{-1}, \\
\]
\[
\delta b(\lambda)c(\lambda)+d(\lambda)d(\lambda)=\delta d(\lambda) \bigg[\frac{a(\lambda)d(\lambda)-1}{d(\lambda)}+\delta^{-1}d(\lambda) \bigg]=\delta d(\lambda)(a(\lambda)+\delta^{-1} d(\lambda))-\delta.
\]

We initially define the renormalization map $\rho: \mathbb{C}^2 \rightarrow \mathbb{C}^2$ in terms of the above diagonal entries as
\[
\rho(x, y)=(x(x+\delta^{-1}y) - \delta^{-1}, \delta y(x+\delta^{-1} y)-\delta)
\]
and the map $\phi:\mathbb{C} \rightarrow \mathbb{C}^2$ as 
\[
\phi(\lambda)=\begin{bmatrix} a(\lambda) \\ d(\lambda) \end{bmatrix}.
\]

\noindent Note that $\rho \circ \phi(\lambda))=\phi(\gamma \lambda)$ for all $\lambda \in \mathbb{C}$.

We now go back to the 2-dimensional complex projective space $\mathbb{P}^2=\mathbb{P}^2(\mathbb{C})$, and note that any point $[x, y,z] \in \mathbb{P}^2$ is equivalent to $[\frac{x}{z}, \frac{y}{z}, 1]$ for $z \neq0$.  We can therefore represent $\mathbb{P}^2$ by 
\[
\mathbb{P}^2= \{(q_1, q_2, 1): (q_1, q_2) \in \mathbb{C}^2\} \cup \{[x, y, 0]: (x, y) \in \mathbb{C}^2\}.
\]

\noindent We can then naturally define the map $\rho$, now viewed as a polynomial map from the complex projective plane $\mathbb{P}^2=\mathbb{P}^2(\mathbb{C})$ to itself, as follows:
\begin{equation}
\label{rho}
\rho([x, y,z])=[x(x+\delta^{-1}y) - \delta^{-1}z^2, \delta y(x+\delta^{-1} y)-\delta z^2, z^2],
\end{equation}
which is in agreement with Equation \eqref{RenormalizationMap}.  

In light of the above discussion, the \emph{invariant curve} $\phi$ can be viewed as a map $\phi: \mathbb{C} \rightarrow \mathbb{C}^3$ defined by $\phi(\lambda)=(a(\lambda), b(\lambda), 1)$ and satisfying the \emph{functional equation}
\begin{equation}
\label{RenormalizationMapIdentity}
\rho \circ \phi(\lambda) = \phi(\gamma \lambda),
\end{equation}
for all $\lambda \in \mathbb{C}$.

Next, we study the spectrum of the eigenvalue equation \eqref{EigenProb}, as well as of its counterpart for $n=\infty$.   
An \textit{attractive fixed point} $x_0$ of $\rho$ is a point such that $\rho x_0 = x_0$ and for any other point $x$ in some neighborhood of $x_0$, the sequence $\{\rho^nx\}_{n=0}^{\infty}$ converges to $x_0$.  The \textit{basin of attraction} of a fixed point is contained in the Fatou set of $\rho$.  For $\delta >1$, $x_0= [0, 1, 0]$ is an attractive fixed point of $\rho$.  The set 
\begin{equation}
\label{SetD}
D := \{[x,y,z]: x+\delta^{-1} y = 0\}
\end{equation}
is part of the Fatou set of $\rho$ since it is contained in the basin of attraction of $x_0$.  (For various notions of higher-dimensional complex dynamics, we refer to the surveys provided in \cite{JF96} and \cite{FS94}.)  The set $D$ and the invariant curve $\phi$ of $\rho$ together determine the spectrum of $H_{<n>}$ and of $H_{<\infty>}$.  Moreover, the set of eigenvalues (i.e., here, the spectrum) can be described by the set 

\begin{equation}
\label{SetS}
S := \{\lambda \in \mathbb{C}: \phi(\gamma^{-1}\lambda) \in D\},
\end{equation}
the `time intersections' of the curve $\phi(\gamma^{-1} \lambda)$ with $D$.  It turns out that $S$ is countably infinite and contained in $(0, \infty)$.    We write $S= \{\lambda_k\}^{\infty}_{k=1}$, with $\lambda_1 \leq \lambda_2 \leq... \leq \lambda_k \leq...$, each eigenvalue being repeated accordingly to its multiplicity.  (It also turns out that the $\lambda_k$'s are all simple, in this case.)  Furthermore, following \cite{LL12}, we call $S$ the \textit{generating set} for the spectrum of $H_{<n>}$, with $n=0,1,...,\infty$. 

Let $S_p = \gamma^p S$, for each $p \in \mathbb{Z}$.  Recall that we are assuming throughout that $\alpha \leq \frac{1}{2}$ (i.e., $\delta \leq 1$).  It follows that not only the spectrum of $H_{<n>}$ (for $n=0, 1, 2,...$) is discrete but unless $\alpha=\frac{1}{2}$ (i.e., $\delta=1$ or equivalently, $\gamma=4$), so is the Dirichlet spectrum of $H_{<\infty>}$.  Furthermore, the spectrum of $H_{<n>}$ and of $H_{<\infty>}$ can be deduced from that of $H_{<0>}$, as will be seen in the next theorem.  Finally, recall that we always have $\gamma \geq 4$ and hence, $\gamma >1$.

\begin{theorem}[Sabot, \text{\cite{CS05}}]
\label{Spectrum}
The spectrum of $H_{<0>}$ on $I =I_{<0>}$ is equal to $\bigcup_{p=0}^{\infty} S_p$, while \emph(if $\alpha < \frac{1}{2}$\emph) the spectrum of $H_{<\infty>}$ on $I_{<\infty>}=[0, \infty)$ is equal to $\bigcup_{p=-\infty}^{\infty}S_p$.\footnote{If $\alpha=\frac{1}{2}$, then the spectrum of $H_{<\infty>}$ is given by the closure of $\cup_{p=-\infty}^{\infty} S_p$.}  Furthermore, for any $n \geq 0$, the spectrum of $H_{<n>}$ is equal to $\bigcup_{p=-n}^{\infty}S_p$.  Moreover, for $n=0, 1,..., \infty$, each eigenvalue of $H_{<n>}$ is simple.  \emph(In particular, each $\lambda_j \in S$ has multiplicity one.\emph)
\end{theorem}

The diagram associated with the set of eigenvalues of the operator $H_{<\infty>}$ can be represented as follows:

\[\begin{array}{ccccc}
\vdots & \vdots & \vdots & \vdots & \\
\gamma^{-2}\lambda_1 & \gamma^{-2}\lambda_2 & \gamma^{-2}\lambda_3 & \gamma^{-2}\lambda_4 & \cdots \\
\gamma^{-1}\lambda_1 & \gamma^{-1}\lambda_2 & \gamma^{-1}\lambda_3 & \gamma^{-1}\lambda_4 & \cdots\\

\lambda_1 & \lambda_2 & \lambda_3 & \lambda_4 & \cdots \\
\gamma\lambda_1 & \gamma\lambda_2 & \gamma\lambda_3 & \gamma\lambda_4 & \cdots \\
\gamma^2\lambda_1 & \gamma^2\lambda_2 & \gamma^2\lambda_3 & \gamma^2\lambda_4 & \cdots\\
\vdots & \vdots & \vdots & \vdots & \\
\end{array}
\]

Sabot's work (\cite{CS98}--\cite{CS05}) has sparked an interest in generalizing the decimation method to a broader class of fractals and therefore, to the iteration of rational functions of several complex variables.  

For each $k \geq 1$, we denote by $f_{k}$ the solution of the equation $H_{<\infty>}f = \lambda_{k} f$ for $\lambda_{k} \in S$.  In other words, $f_k$ is an eigenfunction of $H_{<\infty>}$ associated with the eigenvalue $\lambda_k \in S$.  (Note that $f_k$ is uniquely determined, up to a nonzero multiplicative constant which can be fixed by a suitable normalization.)

\begin{theorem}[Sabot, \text{\cite{CS05}}] \emph{Assume that} $\alpha < \frac{1}{2}.$
\label{EigenValuesFunctions}
\hspace{1mm}
\begin{enumerate}[\emph(i\emph)]

\item Given any $k \geq 1$ and given $p \in \mathbb{Z}$, if $f_{k}$ is the normalized solution of the equation $H_{<\infty>}f = \lambda_{k} f$ for $\lambda_{k} \in S$, then $f_{k,p}:=f_{k}\circ\Psi_1^{-p}$ is the solution of the equation $H_{<\infty>}f = \lambda_{k,p}f$, where $\lambda_{k,p} := \gamma^p\lambda_k$ and $p \in \mathbb{Z}$ is arbitrary.

\item Moreover, if $f_{k,p}= f_{k}\circ\Psi_1^{-p}$ is the solution of the equation $H_{<\infty>}f = \lambda_{k,p}f$, then $f_{k,p,<n>}:=f_{k,p}|_{I_{<n>}}$, the restriction of $f_{k,p}$ to $I_{<n>}$, is the solution of the equation $H_{<n>}f = \lambda_{k,p}f$.
\end{enumerate}

Finally, for each $n=0, 1, ..., \infty$,
$
\{f_{k, p, <n>}: k\geq 1, p \geq -n\}
$
is a complete set of eigenfunctions of $H_{<n>}$ in the Hilbert space $L^2(\mathbb{R}^+, m_{<\infty>})$, where $\mathbb{R}^{+}=I_{<\infty>}=[0, \infty)$.
 \end{theorem}

\section{An infinite lattice based on the Sierpinski gasket}
In this section, we will show that the polynomial induced by the decimation method in the case of the classical (bounded) Sierpinski gasket $SG$ can be recovered from the infinite lattice $SG^{(\infty)}$ based on the (bounded) Sierpinki gasket.

\begin{figure}
  \centering
    \includegraphics[scale=0.47]{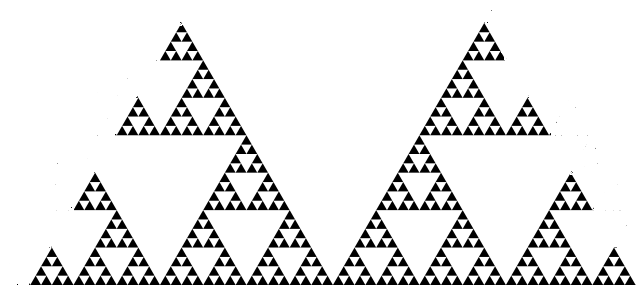}
          \caption{An infinite Sierpinski gasket, $F_{<\infty>}=SG^{(\infty)}$. \label{fig:UnboundedSG}}
\end{figure}

We start with a self-similar set $F=\{1, 2, 3\}$, the vertices of an equilatral triangle, and construct an increasing sequence of finite sets $F_{<n>}$ by blowing-up the initial set $F=F_{<0>}$.  For instance, $F_{<1>}$ is defined as the union of three copies of $F$.  Namely, $F=\cup_{i=1}^{3} F_{<1>, i}$.  The unbounded set $F_{<\infty>}$, called the \emph{infinite Sierpinski gasket} and also denoted by $SG^{(\infty)}$ (see Figure \ref{fig:UnboundedSG}), is the countable set defined as the union of all the finite sets $F_{<n>}=\Phi^{-1}_{i_1} \circ \Phi^{-1}_{i_2} \circ...\circ \Phi^{-1}_{i_n}(F)$, where $(i_1, i_2, ...i_n) \in \{0, 1, 2 \}^n$, $n \geq0$, and $(\Phi_0, \Phi_1, \Phi_2)$ are the contraction mappings, expressing the self-similarity of the set $F$: 
\[
F_{<\infty>}=\bigcup_{n=0}^{\infty} F_{<n>}.  
\]

Similarly, we can define a sequence of operators $H_{<n>}$ on $F_{<n>}$, and the Laplace operator $H_{<\infty>}$ on $F_{<\infty>}$ can be viewed as a suitable limit of the operators $H_{<n>}$.  
The operators $H_{<n>}$ arise from the Dirichlet forms $A_{<n>}$ defined on $\ell(F_{<n>})=\mathbb{R}^{F_{<n>}}$, the space of real-valued functions on $F_{<n>}$.  To construct the Laplacian on the sequence $F_{<n>}$, define the Dirichlet form by 
\[
A_{<n>}(f(x))=\sum_{y \sim x} (f(y)-f(x))^2,
\]
where $f \in \mathbb{R}^{F_{<n>}}$ and the sum runs over all the neighbors $y$ of $x$ in the finite graph associated with $F_{<n>}$.

The operator $H_{<n>}$ on $L^2(F_{<n>}, b_{<n>})$ is the infinitesimal generator of the Dirichlet form defined by $<A_{<n>}f,g> = -\int H_{<n>}f g db_{<n>}$, where $b_{<n>}$ is the positive measure on $F_{<n>}$ which gives a mass of 1 to the points in $\partial F_{<n>}$ and 2 to the points of $F_{<n>}\setminus\partial F_{<n>}$, where $\partial F_{<n>}$ denotes the boundary of the graph $F_{<n>}$.  The sequence $H_{<n>}$ is uniformly bounded and we can use it to define the operator $H_{<\infty>}$ on $L^2(F_{<\infty>},b_{<\infty>})$, where $b_{<\infty>}$ is the positive measure on $F_{<\infty>}$ defined as a suitable limit of the measures $b_{<n>}$.

Sabot \cite{CS03} has shown that the extended decimation method which he established for these fractals naturally involved the complex dynamics of a renormalization map of several complex variables. 

Let $G$ be the group of symmetries acting on the finite lattices $F_{<n>}$, namely $S_3$, which is also the natural symmetry group of the equilateral triangle and of $F=\{1, 2, 3\}$.  We denote by $Sym^G$ the space of symmetric (linear) operators on $\mathbb{C}^F$ which are invariant under $G$.  For every $Q \in Sym^G$, we can construct a symmetric operator $Q_{<1>}$ on $\mathbb{C}^{F_{<1>}}$ as a sum of the copies of $Q$ on $F_{<1>, i}$ (where we recall that $F_{<1>, 1}$, $F_{<1>, 2}$ and $F_{<1>, 3}$ are the three copies of $F_{<1>}$):
\[Q_{<1>}=\sum_{i=1}^3Q_{<1>, i}.
\]

We define a rational map $T: Sym^G \rightarrow Sym^G$ by 
\begin{equation}
\label{RationalMap}
T(Q)=(Q_{<1>})|_{\partial F_{<1>}},
\end{equation}
the trace of $Q$ on the `boundary' of $F_{<1>}$.  In general, using the interpretation of symmetric matrices, it is difficult to analyze the map $T$, which involves the notion of trace on a subset.  (Clearly, one can also view $Q$ as a quadratic form on $\mathbb{C}^F$; hence, the notation.)  To avoid the complication of taking the trace of a symmetric matrix on a subset, the space of symmetric matrices can be embedded into a Grassmann algebra, in which case the map T becomes linear.

Let $\bar{E}$ and $E$ be two linear subspaces of $\mathbb{C}^F$ with canonical basis ${(\bar{\eta}_x)_{x\in F}}$ and $(\eta_x)_{x \in F}$, respectively, and such that $\mathbb{C}^F= \bar{E} \oplus E$.  We define the \emph{Grassmann algebra} associated with this (orthogonal direct sum) decomposition of $\mathbb{C}^F$ by
\[
\bigwedge(\bar{E} \oplus E)=\bigoplus^{2|F|}_{k=0}(\bar{E} \oplus E)^{\wedge k},
\]
where $(\bar{E} \oplus E)^{\wedge k}$ is the k-fold antisymmetric tensor product of $ \bar{E} \oplus E$ with itself, and $|F|=3$ is the cardinality of $F$ .  We consider the subalgebra of the Grassmann algebra generated by the monomials containing the same number of variables $\bar{\eta}$ and $\eta$, namely,
\[
\mathcal{A}=\bigoplus_{k=0}^{|F|} \bar{E}^{\wedge k} \wedge E^{\wedge k}.
\]
We can embed $Sym^G$ into $\mathcal{A}$ via the injection map

\begin{eqnarray*}
Sym^G & \rightarrow & \mathcal{A}
\\ Q &\mapsto & \exp(\bar{\eta}Q \eta),
\end{eqnarray*}
where $\bar{\eta}Q \eta=\sum_{i, j \in F}{Q_{i, j}\bar{\eta}_i\eta_j}$ and $Q:=(Q_{i,j})_{i,j \in F}$, a symmetric matrix.  We denote by $\mathcal{P}(\mathcal{A})$ the projective space associated with $\mathcal{A}$.  Let the corresponding canonical projection be $\pi: \mathcal{A} \rightarrow \mathcal{P}({\mathcal{A}})$.  The closure of the set of points of the form $\pi(\exp(\bar{\eta} Q \eta)$ for $Q \in Sym^G$, denoted by $\mathbb{L}^G$, is a smooth submanifold of $\mathcal{P}(\mathcal{A})$ of dimension $dim(Sym^G)$. (We refer, for example, to \cite{JR02} and \cite{EV03} for an introduction to Grassmann algebras and projective geometry.)  

With $T$ defined by \eqref{RationalMap}, we define the following homogeneous polynomial map of degree 3 (recall that $\eta^2=\bar{\eta}^2=0$):
\begin{eqnarray*}
R:\mathcal{A} &\rightarrow & \mathcal{A}
\\ \exp (\bar{\eta}Q\eta) & \longmapsto & det((Q_{<1>})|_{F_{<1>}\backslash \partial F_{<1>}} \exp(\bar{\eta}TQ\eta).
\end{eqnarray*}
This map induces a map $g$ on the projective space $\mathbb{L}^G=\mathbb{P}^1 \times \mathbb{P}^1$ such that $g^n(\pi(x))=\pi(R^n(x))$ for $x \in \pi^{-1}(\mathbb{L}^G)$ and $n\geq 0$.  (Here, $\mathbb{P}^1=\mathbb{P}^1(\mathbb{C})=\mathbb{C} \cup\{\infty\}$ is the complex projective line, also known as the Riemann sphere.)  Since $\mathbb{L}^G$ is a compactification of $Sym^G$, the map $g:\mathbb{L}^G \rightarrow \mathbb{L}^G$ is an extension of the \emph{trace map} $T:Sym^G \rightarrow Sym^G$ from $Sym^G$ to $\mathbb{L}^G$.

More precisely, note that $\mathbb{C}^F$ can be decomposed into a sum of two irreducible representations: $\mathbb{C}^F= W_0 \oplus W_1$, where $W_0$ is the subspace of constant functions and $W_1$ is its orthogonal complement.  Let $p|_{W_0}$ and $p|_{W_1}$ denote the orthogonal projections of $\mathbb{C}^F$ onto $W_0$ and $W_1$, respectively.  Via the isomorphism between $Sym^G$ and $\mathbb{C}^2$ (recall that $F$ has cardinality $3$), every element $Q \in Sym^G$ can be written as $Q^{u_0, u_1}=u_0p|_{W_0} +u_1p|_{W_1}$, where $(u_0, u_1) \in \mathbb{C}^2$.  Then, the \emph{renormalization map} $T$ is defined by 
\[
T(u_0, u_1)=\bigg(\frac{3u_0u_1}{2u_0+u_1}, \frac{u_1(u_0+u_1)}{5u_1+u_0}\bigg).
\]
This map induces another map on $\mathbb{P}^1$, denoted by  $g$ and given by 
\[
g([z_0: z_1])=[z_0(5z_1+z_0): (2z_0+z_1)(z_0+z_1)].
\]
(Here, $[w_0: w_1]$ denotes the homogeneous coordinates of a generic point of the projective line $\mathbb{P}^1=\mathbb{P}^1(\mathbb{C})$.)  Indeed, upon the substitution $z=\frac{u_0}{u_1}$, the map $T$ gives rise to

\[
\begin{aligned}
g(z) &= \frac{\frac{u_0u_1}{2u_0+u_1}}{\frac{u_1(u_0+u_1}{5u_1+u_0}}
=\frac{u_0u_1}{2u_0+u_1} \frac{5u_1+u_0}{u_1(u_0+u_1)}\\
&= \frac{u_0u_1^2(5+\frac{u_0}{u_1})}{u_1(2\frac{u_0}{u_1}+1)u_1^2(\frac{u_0}{u_1}+1)}\\
&=\frac{z(5+z)}{(2z+1)(z+1)}.
\end{aligned}
\]

\vspace{.8pc}
\noindent The dynamic of the map $g$ plays an essential role in the study of the spectrum of the associated symmetric operators.  After having made the additional change of variable $v=\frac{3z}{1-z}$, we obtain the following equation
\[
\frac{3v(2v+5)}{(3v+3)(2v+3)},
\]
from which we recover the polynomial $p(v)=v(2v+5)$.
This polynomial plays a significant role in the case of the bounded Sierpinski gasket (as shown in \S 2) and was introduced (by completely different methods) in the initial work of Rammal \cite{RR84} and of Rammal and Toulouse \cite{RT83}.  We note that the dynamics of rational maps in higher dimension is hidden in the one-dimensional case of the Sierpinski gasket.    

\begin{remark}
A thorough discussion of the symplectic and supersymmetric aspects of the new methods developed in \cite{CS03} \emph(as well as, implicitly, in \cite{CS98} and \cite{CS01}\emph) to extend to the multi-variable case the classic decimation method, is provided in \cite{CS}.  In particular, beside supersymmetry \emph(which is translated mathematically by the presence of Grassmann algebras and variables\emph), geometric quantization and the associated momentum map in symplectic geometry play an important role in this context.
\end{remark}
 
\begin{remark}
A good review of many of the rigorously established properties of the Laplacian on various realizations of ``the" \emph(deterministic\emph) Sierpinski gasket $SG^{(\infty)}$ can be found in \cite{AT98}, both in the case of discrete and continuous spectra.  \emph(We only discuss the case of discrete spectra in the present paper; see, however, Remark \ref{factorizationformula} below.\emph)  Moreover, in \cite{CS03}, the emphasis is on the study of the spectral properies of random \emph(rather than deterministic\emph) realizations of ``the" infinite \emph(or unbounded\emph) Sierpinski gasket $SG^{(\infty)}$.  
\end{remark}

\section{Factorization of the spectral zeta function}
In this section, we show that the spectral zeta function of the Laplacian defined on a (suitable) finitely ramified self-similar set or on an infinite lattice based on this fractal can be written in terms of the zeta function associated with the renormalization map. We will focus here on the case of the Laplacian on the bounded Sierpinski gasket $SG$ (as in \S2 and \cite{AT07}) or on the infinite Sierpinski gasket $SG^{(\infty)}$ (as in \S4 and \cite{LL12}), as well as on the case of fractal Sturm--Liouville differential operators on the half-line $I_{<\infty>}=[0, \infty)$, viewed as a blow-up of the self-similar interval $I=[0, 1]$ (as in \S3 and \cite{LL12}).

\begin{definition}
\label{SpecZetaFunction}
The \emph{spectral zeta function} of a positive self-adjoint operator $L$ with compact resolvent \emph(and hence, with discrete spectrum\emph) is given \emph(for $Re(s)$ sufficiently large\emph) by 
\begin{equation}
\zeta_L(s) = \sum^{\infty}_{j=1}(\kappa_j)^{-s/2},
\end{equation}
where the positive real numbers $\kappa_j$ are the eigenvalues of the operator written in nonincreasing order and counted according to their multiplicities. 
\end{definition}  

A. Teplyaev (\cite{AT07}, see also \cite{AT04}), motivated by the known identity for fractal strings (see Remark \ref{Remark5.3} below), has studied the spectral zeta function of the Laplacian on SG and, in the process, has explored interesting connections between the spectral zeta function and the iteration of the polynomial induced by the decimation method. 
\begin{theorem}[Teplyaev, \cite{AT07}]
\label{TeplyaevMainThm}
The spectral zeta function of the Laplacian on SG is given by

\begin{equation}
\label{TeplyaevEq}
\zeta_{\Delta_{\mu}}(s) = \zeta_{R,\frac{3}{4}}(s)\frac{5^{-\frac{s}{2}}}{2} \bigg(\frac{1}{1-3 \cdot 5^{-\frac{s}{2}}}+\frac{3}{1-5^{-\frac{s}{2}}}\bigg)+\zeta_{R,\frac{5}{4}}(s)\frac{5^{-s}}{2}\bigg(\frac{3}{1-3 \cdot 5^{-\frac{s}{2}}}-\frac{1}{1-5^{-\frac{s}{2}}}\bigg),
\end{equation}
where $R(z):=z(5-4z)$ and 
\begin{equation}
\label{PolynomialZetaFunction}
\zeta_{R,z_0} (s): = \lim_{n\rightarrow \infty}\sum_{z\in R^{-n}\{z_0\}}(c^n z)^{-\frac{s}{2}}
\end{equation}
 is the \emph{polynomial zeta function} of $R$, defined for $Re(s) > \frac{2\log 2}{\log c}$ \emph(where $c:=5=R^{\prime}(0)$\emph).  Furthermore, there exists $\epsilon >0$ such that $\zeta_{\Delta_{\mu}}(s)$ has a meromorphic continuation for $Re(s)>-\epsilon$, with poles contained in $\bigg\{\frac{2in\pi}{\log 5}, \frac{\log 9+2in\pi}{\log 5}: n\in \mathbb{Z}\bigg\}$.
\end{theorem}

\begin{remark}
\label{Remark5.3}
Theorem \ref{TeplyaevMainThm} extends to the present setting of analysis on certain fractals \cite{Kigami01} \emph(or `drums with fractals membrane', see, e.g., \cite{B79}, \cite{DGV08}, \cite{UF03}--\cite{FS92}, \cite{HHW87}, \cite{Kigam93}--\cite{LL12}, \cite{Lapidus93}--\cite{Lapidus95}, \cite{LvanF04}, \cite{RR84}--\cite{CS03}, \cite{TS96}, \cite{RS06}, \cite{AT98}--\cite{AT07}\emph) and provides a dynamical interpretation of the factorization formula obtained by the second author in \cite{Lapidus92}, \cite{Lapidus93} for the spectral zeta function $\zeta_{\nu}(s)$ of the Dirichlet Laplacian associated with a fractal string \emph(a one-dimensional drum with fractal boundary \cite{Lapidus91}--\cite{LvanF06}\emph)\emph:
\begin{equation}
\label{LapidusEq}
\zeta_{\nu}(s)=\zeta(s)\cdot \zeta_g(s),
\end{equation}
where $\zeta(s)$ is the classic Riemann zeta function and $\zeta_g(s)$ is \emph(the meromorphic continuation of\emph) the geometric zeta function of the fractal string \emph(whose poles are called the \emph{complex dimensions} of the string and help describe the oscillations intrinsic to the geometry and the spectrum of the string, see \cite{LvanF06}\emph).  See also \emph(\cite{LvanF06}, \emph{\S1.3}\emph) for a discussion of Equation \eqref{LapidusEq} and, for example, \cite{Lapidus92}, \cite{Lapidus93}, \cite{LM95}, \cite{LP93}, along with much of \cite{LvanF06}, for various applications of this factorization formula.  Finally, we note that in \cite{AT07}, $\zeta(s)$ is reinterpreted as the polynomial zeta function of a certain quadratic polynomial of a single complex variable and hence, the factorization formula \eqref{LapidusEq} can also be  given a complex dynamical interpretation \emph(in terms of the iteration of the renormalization map\emph).  This issue was revisited in \cite{LL12}, and, as a result, formula \eqref{LapidusEq} can also be interpreted in terms of multi-variable complex dynamics.  \emph(See \cite{LL12}, along with Theorem \ref{RiemannZetaFunction} below.\emph)
\end{remark}

\begin{remark}
\label{Remark5.4}
In \cite{DGV08}, shortly after the completion of \cite{AT04}, \cite{AT07}, G. Derfel, P. Grabner and F. Vogl, working independently on this question and motivated in part by the results and conjectures of \cite{Kigam93} and \cite{Lapidus93}, have obtained another interpretation of the geometric factor of the factorization formulas \eqref{TeplyaevEq} and \eqref{LapidusEq}, expressed in terms of the multiplicities of the eigenvalues.  In the process, they have shown that the spectral zeta function $\zeta_{\Delta_{\mu}}(s)$ has a meromorphic continuation to all of $\mathbb{C}$.  Moreover, they have proved further cases of a conjecture of \cite{Kigam93}, \cite{Lapidus93}, according to which the asymptotic second term in the spectral counting function of the Laplacian on the Sierpinski gasket $SG$ and other lattice self-similar fractals is truly oscillatory \emph(or equivalently, the corresponding periodic function obtained in \cite{Kigam93} is not constant\emph).  \emph[Compare with analogous conjectures and results stated or obtained in \cite{Lapidus91}--\cite{LvanF06} for drums with fractal boundary \emph(instead of drums with fractal membrane\emph).\emph]
\end{remark}

Later on, in \cite{LL12}, we extended this result about the factorization of the spectral function of the Laplacian on the finite (or bounded) gasket $SG$ to the infinite (or unbounded) Sierpinski gasket $SG^{(\infty)}$ and to the renormalization maps of several complex variables associated with fractal Sturm--Liouville operators.

The Dirac delta hyperfunction on the unit circle $\mathbb{T}$ is defined as $\delta_{\mathbb{T}}=[\delta^+_{\mathbb{T}}, \delta^-_{\mathbb{T}}]=[\frac{1}{1-z}, \frac{1}{z-1}]$.  It consists of two analytic functions, $\delta^+_{\mathbb{T}}: E \rightarrow \mathbb{C}$ and $\delta^-_{\mathbb{T}}: \mathbb{C} \backslash \bar{E} \rightarrow \mathbb{C}$, where $E=\{z\in \mathbb{C}: |z| <1 + \frac{1}{N}\}$ for a large natural number $N$.  In other words, a hyperfunction on $\mathbb{T}$ can be viewed as a suitable pair of holomorphic functions, one on the unit disk $|z| <1$, and one on its exterior, $|z|>1$.
(See, for example, (\cite{UG10}, \S 1.3) and (\cite{MM76}, \S3.3.2) for a discussion of various changes of variables in a hyperfunction.  See also those two books \cite{UG10}, \cite{MM76}, along with \cite{MS58}, \cite{MS59}, the original articles by M. Sato, for an overview of the theory of hyperfunctions.  Moreover, see \cite{YT87} for a detailed discussion of $\delta_{\mathbb{T}}$ and, more generally, of hyperfunctions on the unit circle $\mathbb{T}$.)

\begin{theorem}[\cite{LL12}, Lal and Lapidus]
\label{InfiniteSGThm}
The spectral zeta function $ \zeta_{\Delta^{(\infty)}}$ of the Laplacian $\Delta^{(\infty)}$ on the infinite Sierpinski gasket $SG^{(\infty)}$ is given by 
\begin{equation}
\label{infiniteSG}
\zeta_{\Delta^{(\infty)}}(s)=\zeta_{\Delta_{\mu}}(s)\delta_{\mathbb{T}}(5^{-\frac{s}{2}}) ,
\end{equation}
where $\delta_{\mathbb{T}}$ is the Dirac hyperfunction on the unit circle $\mathbb{T}$ and $\zeta_{\Delta_{\mu}}$ is the spectral zeta function of the Laplacian on the finite \emph(i.e., bounded\emph) Sierpinski gasket SG, as given and factorized explicitly in Equation \eqref{TeplyaevEq} of Theorem \ref{TeplyaevMainThm}.
\end{theorem}

In the case of the Sturm--Liouville operator on the half-line, we introduce a multi-variable analog of the polynomial zeta function occurring in Equation \eqref{PolynomialZetaFunction} of Theorem \ref{TeplyaevMainThm}.

\begin{definition}[\cite{LL12}]
\label{MultiPolyZeta}
We define the \emph{zeta function of the renormalization map} $\rho$ to be
\begin{equation}
\label{MultiPoly}
\zeta_{\rho}(s) = \sum_{p=0}^{\infty}\sum_{\{\lambda \in \mathbb{C}:\hspace{.5mm } \rho^p(\phi(\gamma^{-(p+1)}\lambda)) \in D\}} (\gamma^p \lambda)^{-\frac{s}{2}},
\end{equation}
for Re\emph(s\emph) sufficiently large.
\end{definition}

\begin{remark}
Recall from \emph{\S3} that in the present situation of fractal Sturm--Liouville operators, the renormalization map $\rho$ is given by Equation \eqref{RenormalizationMap} and that the `renormalization constant' \emph(or `scaling factor'\emph) $\gamma$ is given by Equation \eqref{constants}.  \emph(Also, see the functional equation \eqref{RenormalizationMapIdentity} defining the invariant curve $\phi$, as well as Equations \eqref{SetD} and \eqref{SetS} defining $D$ and the generating set $S$, respectively.\emph)  Furthermore, recall from the end of \emph{\S3.2} that we assume that $\alpha \leq \frac{1}{2}$ \emph(i.e., $\delta \leq 1$\emph) in order to ensure the discreteness of all the spectra involved,\footnote{Except when $\alpha=\frac{1}{2}$, in which case the spectrum of $H_{<\infty>}$ is continuous.} and that we always have $\gamma \geq 4$\emph; in particular, $\gamma>1$, and $\gamma=4$ if and only if $\alpha=\frac{1}{2}$.

\end{remark}

We consider the factorization formulas associated with the spectral zeta functions of the sequence of operators $H_{<n>}=-\frac{d}{dm_{<n>}}\frac{d}{dx},$ starting with $H_{<0>}$  on $[0, 1]$, which converges to the Sturm--Liouville operator $H_{<\infty>}$ on $[0, \infty)$.  In light of Definition \ref{SpecZetaFunction} and Theorem \ref{Spectrum}, given any integer $n\geq 0$, the spectral zeta $\zeta_{H_{<n>}}(s)$ of $H_{<n>}$ on $[0, \alpha^{-n}]$ is initially given by $\zeta_{H_{<n>}}(s) = \sum_{\lambda \in S} \sum_{p=-n}^{\infty} (\gamma^p\lambda)^{-\frac{s}{2}}$, for $Re(s)$ large enough. 

\begin{theorem}[\cite{LL12}]
\label{H0MultiPolyZeta}
The zeta function $\zeta_{\rho}(s)$ of the renormalization map $\rho$ is equal to the spectral zeta function of $H_{<0>}$,
\[
\zeta_{H_{<0>}}(s) = \sum_{\lambda \in S} \sum^{\infty}_{p=0} (\gamma^p\lambda)^{-\frac{s}{2}},
\]
or its meromorphic continuation thereof\emph: $\zeta_{\rho}(s)=\zeta_{H_{<0>}}(s)$.  \emph(An expression for $\zeta_{H_{<0>}}(s)$ is given by the $n=0$ case of Proposition \ref{HnZetaFunction} just below\emph; see Equation \eqref{zetarho} of Remark \ref{PolyGeomEq}.\emph)  \end{theorem}

\begin{proposition}[\cite{LL12}]
\label{HnZetaFunction}
For $n \geq 0$ and $Re(s)$ sufficiently large, we have
\begin{equation}
\label{HnZeta}
\zeta_{H_{<n>}}(s)=\frac{(\gamma^n)^{\frac{s}{2}}}{1-\gamma^{-\frac{s}{2}}} \zeta_S(s),
\end{equation}
where $\zeta_S(s)$ is the \emph{geometric zeta function of the generating set} $S$.  Namely, $\zeta_S(s):= \sum_{j=1}^{\infty}(\lambda_j)^{-\frac{s}{2}}$ \emph{(}for Re\emph{(}s\emph{)} large enough\emph{)} or is given by its meromorphic continuation thereof.
\end{proposition}

\begin{remark}
\label{PolyGeomEq}
In particular, in light of Theorem \ref{H0MultiPolyZeta}, we deduce from  the $n=0$ case of Proposition \ref{HnZetaFunction} that 
\begin{equation}
\label{zetarho}
\zeta_{\rho}(s)=\zeta_{H_{<0>}}(s)=\frac{1}{1-\gamma^{-\frac{s}{2}}} \zeta_S(s).
\end{equation}

\end{remark}

In the case of the operator $H_{<\infty>}$, the asymptotic behavior of the spectrum led us naturally to using the notion of delta hyperfunction.   Indeed, according to Theorem \ref{Spectrum}, a part of the spectrum of $H_{<\infty>}$ tends to 0 while another part tends to $\infty$.  If one mechanically applies Definition \ref{SpecZetaFunction}, one then deduces that $\zeta_{H_{<\infty>}}(s)$ is identically equal to zero, which is clearly meaningless.  Indeed, the geometric factor in the factorization of $\zeta_{H_{<\infty>}}(s)$ is equal to the sum of two geometric series (converging for $Re(s) > 0$ and for $Re(s) < 0$, respectively), and this sum is itself identically equal to zero (except for the fact that one of the two terms in the sum is not well defined, no matter which value of $s \in \mathbb{C}$ one considers).  Fortunately, there is a satisfactory resolution to this apparent paradox.  More specifically, we have discovered that the geometric part of the product formula for the spectral zeta function $\zeta_{H_{<\infty>}}$ can be expressed in terms of the Dirac delta hyperfunction $\delta_{\mathbb{T}}$ on the unit circle.

\begin{theorem}[\cite{LL12}]
\label{HZetaFunction}
Assume that $\alpha < \frac{1}{2}$.  The spectral zeta function $\zeta_{H_{<\infty>}}$ is factorized as follows\emph{:}
\begin{equation}
\label{HZeta}
\zeta_{H_{<\infty>}}(s)= \zeta_S(s) \cdot \delta_{\mathbb{T}}(\gamma^{-\frac{s}{2}}) =\zeta_{\rho}(s) (1-\gamma^{-\frac{s}{2}})\delta_{\mathbb{T}}(\gamma^{-\frac{s}{2}}). 
\end{equation}
\end{theorem}

Furthermore, we have shown in \cite{LL12} that the zeta function associated with the renormalization map coincides with the Riemann zeta function $\zeta(s)$ for a special value of $\alpha$.  When $\alpha= \frac{1}{2}$ (or equivalently, $\delta=1$ and so $\gamma=4$), the self-similar measure $m$ coincides with Lebesgue measure on $[0,1]$ and hence, the `free Hamiltonian' $H=H_{<0>}$ coincides with the usual Dirichlet Laplacian on the unit interval $[0, 1]$. 

Recall that $\zeta_{\rho}(s)=\zeta_{H_{<0>}}(s)$ (by Theorem \ref{H0MultiPolyZeta}) and that the spectrum of $H_{<0>}$ (i.e., of the Dirichlet Laplacian on $[0, 1]$) is discrete and given by $\pi^2 j^2$, for $j=1, 2, ...$.  Hence, $\zeta_{\rho}(s)=\zeta_{H_{<0>}}(s)=\pi^{-s}\zeta(s)$, where $\zeta(s)$ is the Riemann zeta function.

\begin{theorem}[\cite{LL12}]
\label{RiemannZetaFunction}
When $\alpha=\frac{1}{2}$, the Riemann zeta function $\zeta$ is equal \emph(up to a trivial factor\emph) to the zeta function $\zeta_{\rho}$ associated with the renormalization map $\rho$ on $\mathbb{P}^2(\mathbb{C})$.  More specifically, we have 
\begin{equation}
\label{RiemannZeta}
\zeta(s) =\pi^s \zeta_{\rho}(s)= \frac{\pi^s}{1-2^{-s}} \mbox{ }\zeta_S(s),
\end{equation}
where $\zeta_{\rho}$ is given by Definition \ref{MultiPolyZeta} \emph(or its mermorphic continuation thereof\emph) and the polynomial map $\rho: \mathbb{P}^2(\mathbb{C}) \rightarrow \mathbb{P}^2(\mathbb{C})$ is given by Equation \eqref{RenormalizationMap} with $\alpha=\frac{1}{2}$ \emph(and hence, in light of \eqref{constants}, with $\delta=1$ and $\gamma=4$\emph)\emph:
\begin{equation}
\label{RenormalMap}
\rho([x,y,z])= [x(x+y)-z^2, y(x+y)-z^2, z^2].
\end{equation}

\end{theorem}
This is an extension to several complex variables of A. Teplyaev's result \cite{AT07}, which states that the Riemann zeta function can be described in terms of the zeta function of a quadratic polynomial of one complex variable (as defined by Equation \eqref{PolynomialZetaFunction}).

\begin{remark}
\label{openproblem}
An interesting open problem is to determine what happens if we consider $\zeta_{H_{<\infty>}}(s)$ instead of $\zeta_{H_{<0>}}(s)$, still when $\alpha=\frac{1}{2}$.  In that case, the spectrum of $H_{<\infty>}$ is continuous and a suitable interpretation has to be found for $\zeta_{H_{<\infty>}}(s)$, in terms of a properly defined spectral density of states.  We leave the investigation of this open problem for a future work.  \emph(See also Remark \ref{factorizationformula} just below.\emph)

\end{remark}

\begin{remark}
\label{factorizationformula}
It would be interesting to obtain similar factorization formulas for more general classes of self-similar fractals and fractal lattices, both in the present case of purely discrete spectra or in the mathematically even more challenging case of purely continuous \emph(or mixed continuous and discrete\emph) spectra.  The latter situation will require an appropriate use of the notion of density of states, of frequent use in condensed matter physics \emph(see, e.g., \cite{AO82}, \cite{NY03}, \cite{AO96}\emph) and briefly discussed or used mathematically in various related settings \emph(involving either discrete or continuous spectra\emph) in, e.g.,  \cite{FS92}, \cite{HHW87}, \cite{Kigam93} and especially, \cite{CS98}, \cite{CS01}, \cite{CS03}.
\end{remark}

\bibliographystyle{amsalpha}

\end{document}